\documentclass[twocolumn,secnumarabic,amssymb,nobibnotes,aps,superscriptaddress, prl, reprint]{revtex4-1}
\usepackage{graphicx}
\graphicspath{{figures/}} 
\usepackage{bm}
\usepackage{float}
\usepackage{braket}
\usepackage{upgreek}
\usepackage{amsmath}

\begin{document}

\title{Thermalization of nuclear spins in lanthanide molecular magnets}

\author{Gheorghe Taran}%
\email[]{gheorghe.taran@kit.edu}
\author{Edgar Bonet}%
\email[]{bonet@grenoble.cnrs.fr}
\affiliation{Univ. Grenoble Alpes, CNRS, Grenoble INP\footnote{Institute of Engineering Univ. Grenoble Alpes}, Institut N\'eel, 38000 Grenoble, France}
\author{Wolfgang Wernsdorfer}%
\email[]{wolfgang.wernsdorfer@kit.edu}
\affiliation{Physikalisches Institute, KIT, Wolfgang-Gaede-Str. 1, Karlsruhe D-76131}
\affiliation{Univ. Grenoble Alpes, CNRS, Grenoble INP\footnote{Institute of Engineering Univ. Grenoble Alpes}, Institut N\'eel, 38000 Grenoble, France}
\affiliation{Institute of Nanotechnology (INT), Karlsruhe Institute of Technology (KIT), Hermann-von-Helmholtz-Platz 1, D-76344 Eggenstein-Leopoldshafen}

\date{\today}%

\begin{abstract}
Single molecule magnets distinguish themselves in the field of quantum magnetism through the ability to combine fundamental research with promising applications, the evolution of quantum spintronics in the last decade exemplifying the potential held by molecular based quantum devices.
Notably, the read-out and manipulation of the embedded nuclear spin states was used in proof of principle studies of quantum computation at the single molecule level.
In this paper we study the relaxation dynamics of the $^{159}$Tb nuclear spins in a diluted molecular crystal by using recently acquired understanding of the nonadiabatic dynamics of TbPc$_2$ molecules.
We find that phonon modulated hyperfine interaction opens a direct relaxation channel between the nuclear spins and the phonon bath. 
We highlight the potential importance of the discovered mechanism for the theory of spin bath and the relaxation dynamics of the molecular spins at crossover temperatures.
\end{abstract}

\maketitle

The advancement towards industrially viable quantum technologies like quantum computing and nanoscale magnetometry depends largely on our ability to control the immediate environment of a system of interacting quantum objects (qubits).
The main objective is to preserve coherence during external manipulations and thus exploit intrinsic quantum properties like interference and entanglement.
Depending on their nature and coupling strength, most environmental degrees of freedom in interaction with the qubit can be mapped either into a bosonic bath, for non-local, weak interaction, or to a spin bath, in the case of localized, strong interactions~\cite{prokof2000theory}.
The later case is especially important as it can induce decoherence even in the $T \xrightarrow~ 0$ limit, that is, the dephasing is not accompanied by dissipative processes.
Thus, the complex problem of a central quantum system coupled to localized environmental excitations, for example, nuclear spins or paramagnetic centers, is pivotal in mesoscopic quantum physics~\cite{yang2016quantum}.
Amongst experimental implementations of the above model (\textit{e.g}, NV centers in diamond~\cite{awschalom2007diamond}, nanomagnets~\cite{morello2008quantum}, SQUIDs~\cite{tomsovic1998tunneling} and impurities in silicon~\cite{awschalom2007diamond}), molecular magnets illustrate especially well the intimate relationship that exists between the manifested quantum phenomenologies (\textit{e.g.} quantum tunneling of magnetization (QTM)~\cite{friedman1996macroscopic}, spin parity effect~\cite{wernsdorfer2002spin}, Rabi oscillations~\cite{bertaina2008quantum}) and the spin bath. 
After the experimental evidence of the magnetic bistability in Mn$_{12}$-ac~\cite{sessoli1993}, event that marks the birth of the field of molecular magnetism, the breakthrough discoveries of both phonon-assisted and ground state quantum tunneling~\cite{friedman1996,sangregorio1997} greatly boosted the interest in these systems.
Theoretical inquires to explain the observed dynamics were resolved by carefully considering the effect of the environmental interactions (both spin-phonon and spin-spin couplings) and thus contributed significantly to the development of the theory of the spin bath~\cite{prokof1998low}.
Molecular magnets also proved to be ideal systems for testing the predictions made by the constructed theory.
Thus, both the influence of the isotopic composition through variation of the hyperfine interaction on the relaxation rate~\cite{wernsdorfer2000effects} and the peculiar square root law for the relaxation at low temperatures and short times were promptly verified~\cite{wernsdorfer1999nuclear}.
The strong correlations between the dynamics of the spin bath and the relaxation of the molecular spin were also evidenced by measuring directly the nuclear spins through resonant techniques~\cite{furukawa2001magnetic}.
Both the longitudinal and transverse relaxation times were linked directly to the electronic spin dynamics~\cite{morello2004nuclear} proving that nuclear spins can serve as microscopic probes for the molecular spin~\cite{jang2000measurement}.
The theory of the spin bath was also successfully applied in the study of the decoherence in crystals of molecular magnets~\cite{takahashi2009coherent,takahashi2011decoherence}, paving the way for molecular optimization for quantum information processing nanodevices.

Most of the experimental and theoretical investigations into the subject of spin-bath were done using transition metal ion compounds as model systems (\textit{e.g.} Mn$_{12}$-ac and Fe$_8$) because these were the first discovered single molecule magnets (SMMs) and for a long time remained the best understood compounds~\cite{gatteschi2003quantum}. 
However, the last decade saw the rise of molecular complexes that employ lanthanide ions as magnetic centers and at the moment one can make an argument that this class of SMMs are amongst the most promising ones.
Recent achievements include the observation of magnetic bistability of a Dy complex at temperatures well above liquid nitrogen ~\cite{goodwin2017molecular} and the implementation of the quantum Grover algorithm at the single molecule level~\cite{godfrin2017operating}.

Unquenched orbital angular momentum, large single ion anisotropy and a strong hyperfine interaction are just some characteristics that distinguish lanthanide complexes in the field of molecular magnetism~\cite{woodruff2013lanthanide}.
For example, it was shown that the nature of the strong interaction between the electronic shell of the lanthanide ion and it’s own nuclear spin has strong repercussions on the tunneling dynamics~\cite{ishikawa2005quantum,taran2019role}.
Also, the non-zero orbital momentum brings upfront the spin-phonon interaction. 
Thus, even in the range where the dynamics is temperature independent, the phonon bath can no longer be ignored.
Despite these observations, the quantum dynamics of lanthanide SMMs in the framework of the spin bath theory is a subject largely unexplored.

In the theoretical endeavour that one has to undergo in order to improve our understanding of the displayed dynamics, the properties of the lanthanide molecular complex immersed in a bath of both bosonic and spin nature has to be considered. 
The main difficulties arise when the dynamics of the bath is strongly coupled to the dynamics of the central spin as the effect of the environment cannot be treated perturbatively ~\cite{morello2008quantum}.
However, the analysis of the properties of the spin bath when the molecular spins are static is an important first step.
 
In this paper we try to solve a piece of the puzzle and investigate the thermalization of the $^{159}$Tb nuclear spins in a crystal of prototypical TbPc$_2$ molecules. 
We chose the TbPc$_2$ complex as a model system because we can use the recently acquired understanding on the dynamics of its magnetization to develop a read-out technique for the population of the hyperfine states.
Thus, we will start by analyzing the TbPc$_2$'s sub-kelvin magnetization dynamics in a varying magnetic field and its link to the distribution of the nuclear spins.
Then, the time evolution of the population of the hyperfine states, obtained by fitting the magnetization curves, is evaluated in the framework of a Markovian master equation that allows us to discuss the dynamics in terms of spin-phonon relaxation rates. 
Finally, by evaluating the temperature dependence of the relaxation process we identify the main mechanism responsible for the thermalization of the nuclear spins.
We find that a direct process that involves phonon modulation of the hyperfine interaction is sufficient to explain the magnitude of the determined relaxation rates.

\begin{figure}
\includegraphics[width=0.5\textwidth]{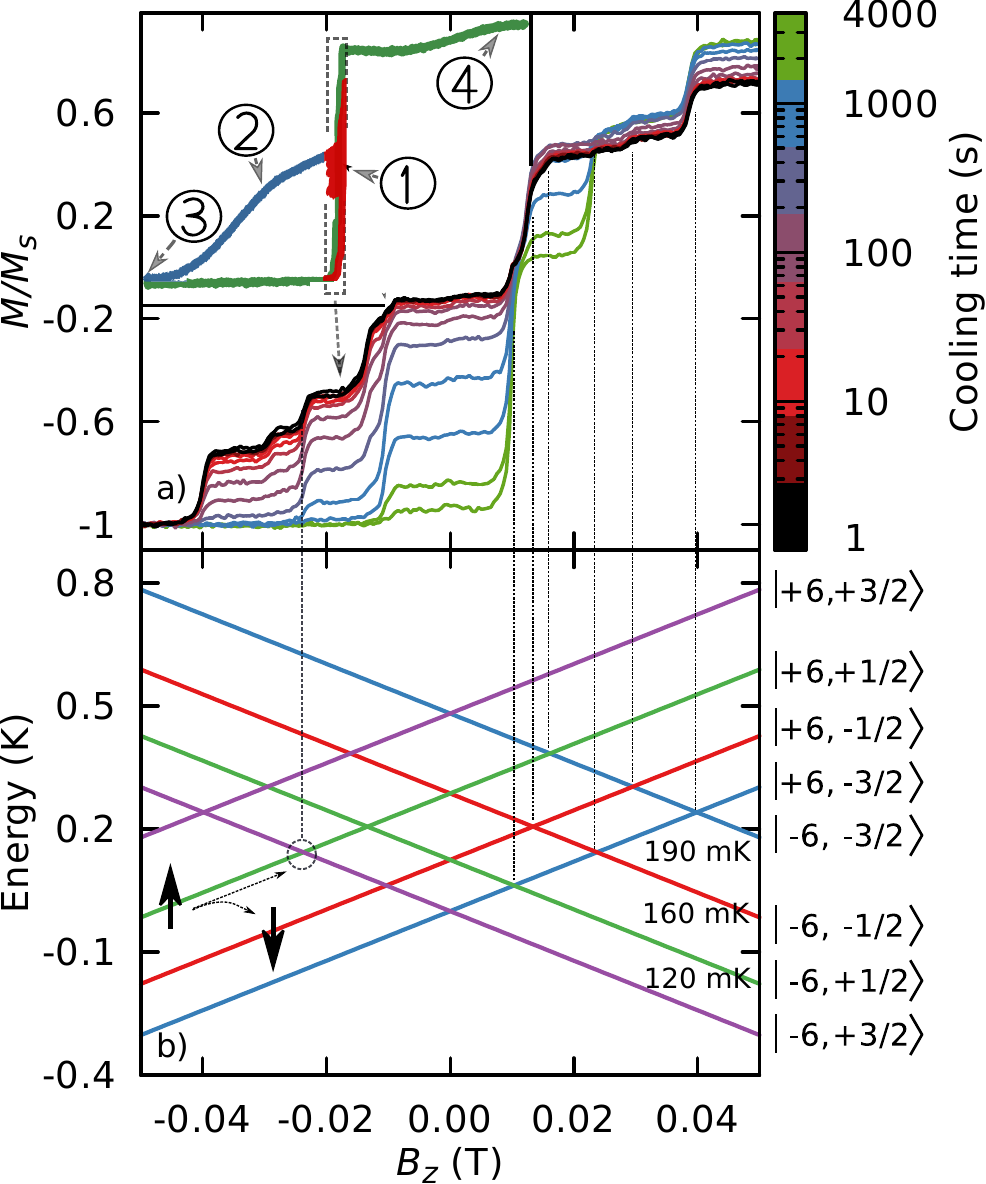}
    \caption{
    \label{fig:therm_zd}
    (a) Magnetization curves of a diluted TbPc$_2$ crystal measured with the $\upmu$SQUID technique at 50~mK as a function of the time the sample is kept in $B_z = -1.3$~T (cooling time). The full procedure to measure these magnetization curves is show in the inset: (1) initialize the sample by sweeping through the zero field resonances multiple times until the $M = 0$ state is reached, (2) saturate the sample, (3) wait a certain time for nuclear spins to thermalize and (4) read-out the hyperfine populations by inverting the magnetization while measuring $M(B_z)$.
(b) Hyperfine structure of the ground doublet, $m_J = \pm 6$ , as a function of the applied longitudinal field obtained after numerical diagonalization of the Hamiltonian given by Eq.~(\ref{eq:H}).
The large black arrows are meant to illustrate how the  $^{159}$Tb transitions from the state $\ket{+6, +1/2}$ to the state $\ket{+6, -1/2}$ during the thermalization process are seen in the magnetization curve as a decrease in the magnitude of the relaxation step.
    	}
	\vspace{-15pt}
\end{figure}

\vfill
\noindent
\textbf{Results}\\
\textbf{TbPc$_2$ single-molecule magnet.}
We report $\upmu$SQUID measurements on micrometer sized crystals containing TbPc$_2$ molecules (Fig.~\ref{fig:therm_zd}a) that are diluted in an isostructural, diamagnetic matrix formed by YPc$_2$ molecules with a concentration of about 1\% (see Methods).
The sample's dilution controls the dipolar interaction between the molecular spins and consequently is used to reduce the probability of the collective effects.
Thus, on one hand, a diluted sample shows a well resolved hyperfine structure and on the other, its dynamics can be understood in terms of the properties of an ensemble of non-interacting molecular spins.

The TbPc$_2$ molecule features a Tb$^{3+}$ ion sandwiched between two phthalocyanine planes in a square antiprismatic symmetry (D$_{4d}$).
The magnetic properties of the compound are dominated by the single ion anisotropy of the Tb$^{3+}$ ion that gives a spin ground state $J = 6$ and by the uniaxial character  of the ligand field interaction that further splits the $2J+1$ degenerate eigenstates.
As a result of the above interactions, in zero external field, we obtain a ground state doublet ($m_J = \pm 6$) that is separated from the first excited doublet ($m_J' = \pm 5$) by about 600~K~\cite{ishikawa2003lanthanide}.
Thus, at temperatures much lower that the zero field splitting, the molecular spin $J = 6$ can be described as an effective spin 1/2 with an effective g-value, $g_{\text{eff}} = 18$.
Under an external field parallel to the anisotropy z-axis, the effective two level Hamiltonian describing the electronic states can be written as: 
${\cal H}_{\text{e}} = \frac {g_{\text{eff}}}{2} \mu_{\text{B}}B_z \sigma_z + \frac {\Delta}{2} \sigma_x$. 
Where the first term represents the longitudinal Zeeman interaction and the second one models the non-axial ligand field interactions. 
The tunnel splitting, $\Delta$, was shown to be in the $\upmu$K range~\cite{taran2019decoherence}.
    
The $^{159}$Tb nucleus that lies at the heart of the molecule also has a non-zero nuclear spin, $I=3/2$, which couples to the surrounding electronic shell and further splits the ground doublet, $m_J = \pm 6$ in a manifold of four levels.
The interaction is modelled by adding a hyperfine and a nuclear quadrupolar contribution to the spin Hamiltonian. 
Thus, the total Hamiltonian is now:     
\begin{align}
\label{eq:H}
\mathcal{H}_{\text{TbPc$_{2}$}}  
=  \mathcal{H}_{\text{e}} +  A_{\text{hyp}} |m_J|({\bm\sigma} \cdot \mathbf{I}) +
\mathbf{I}\hat{P}_{\text{quad}}\mathbf{I}
\end{align}
The isotropic hyperfine interaction, $A_{\text{hyp}} |m_J|({\bm\sigma} \cdot \mathbf{I})$, has three components: the Fermi contact interaction, the paramagnetic spin-orbit contribution and the dipole-dipole interaction resulting in the hyperfine constant $A_{\text{hyp}} = 26.7$~mK.
Because $I > 1/2$, the $^{159}$Tb nucleus has a quadrupolar moment that couples to the electric field gradient through $\mathbf{I}\hat{P}_{\text{quad}}\mathbf{I}$ with the dominant term being the axial component with magnitude $P_{\text{quad}} = 17$~mK.
The tensorial nature of $\hat{P}_{\text{quad}}$ reflects the non-axial character (with respect to the easy axis of the ligand field) of the quadrupolar interaction which together with $\frac {\Delta}{2} \sigma_x$ of the electronic Hamiltonian was shown to be responsible for the observed tunneling dynamics of TbPc$_2$ in diluted single crystals~\cite{taran2019role}.

We label the states by using the electronic and nuclear spin components, $\ket {m_J, m_I}$, with $m_J = \pm 6$ and $m_I = -3/2 \dots 3/2$. 
We write $\mathcal{H}_{\text{TbPc$_{2}$}}$ numerically in the above basis. As we diagonalize it for a magnetic field in the range $\left[-50:50\right]$~mT, we obtain the Zeeman diagram shown in Fig.~\ref{fig:therm_zd}b.
We can see that, the interaction between the electronic spin and nuclear spin results in the non-equidistant splitting of the energy levels, with the energy spacings between consecutive hyperfine states being around 120, 160 and 190~mK.
For temperatures comparable to the hyperfine splitting, a non-equilibrium distribution of the population of the hyperfine states is expected to evolve towards the Boltzmann distribution on a time scale determined by the relaxation mechanism(s).

\vfill
\noindent
\textbf{Thermalization of $^{159}$Tb nuclear spins.}
In what follows, we will first show the experimental evidence of the thermalization process and then present the protocol and theoretical tools used to describe the relaxation process.

As already mentioned, in the present study we use $\upmu$SQUID magnetometry which is sensitive to the magnetic field generated by the molecular spins moments.
To understand qualitatively the link between the magnetization curve ($M(B_z)$) exhibited by the TbPc$_2$'s diluted crystal and the population of the hyperfine states let's look at the $M(B_z)$ characteristic at 50~mK (Fig.~\ref{fig:therm_zd}a and Ref.~\cite{ishikawa2005quantum} for more details).

If we start with a saturated sample in a large magnetic field applied along the easy axis ($B_z \approx -1.3$~T) and sweep the magnetic field, then no relaxation is observed until we reach the level crossings in the Zeeman diagram (Fig.~\ref{fig:therm_zd}b).
This ascertains that the temperature is low enough so that the over-barrier relaxation through the interaction with the lattice vibrations is an improbable process.
As we reach the first crossing in the Zeeman diagram ($B_z \approx -40$~mT) a relaxation step in the magnetization curve is observed.
This bears evidence to resonant quantum tunneling processes induced by the non-axial interactions. 
The height of the relaxation step depends both on the tunneling probability and on the population of the levels that form the anticrossing.
Thus, if the tunneling probability is known we can obtain the population of the hyperfine states by fitting the magnetization curve.

\begin{figure}
    \centering
    \begin{minipage}{0.5\textwidth}
        \includegraphics[width=\textwidth]{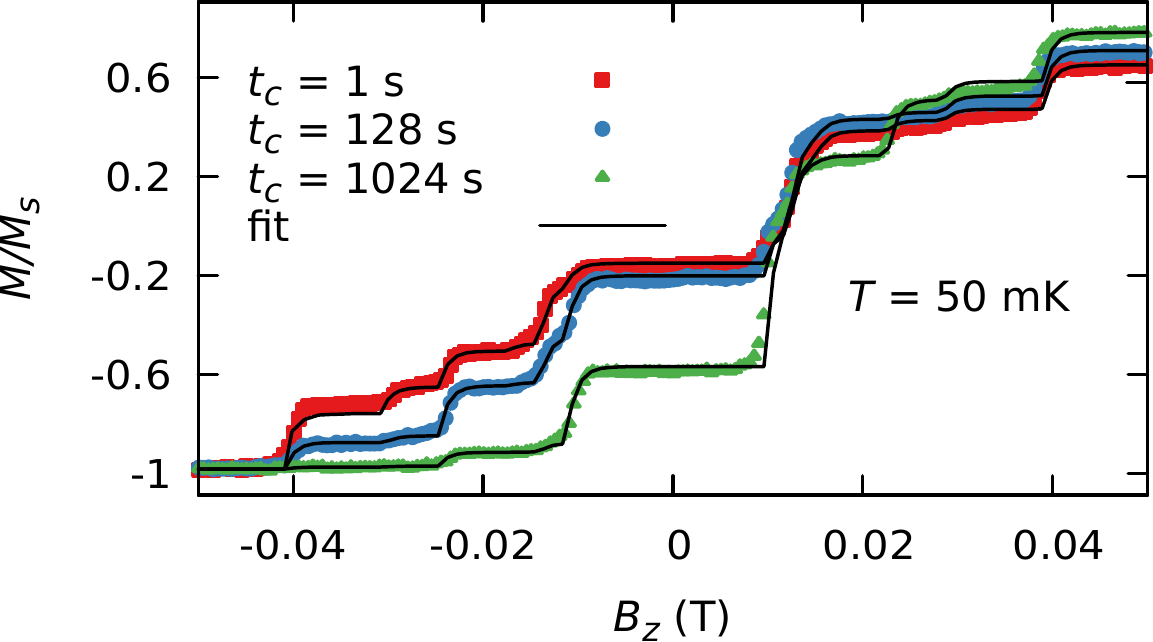}
    \end{minipage}
    \vfill
    \begin{minipage}{0.5\textwidth}
        \includegraphics[width=\textwidth]{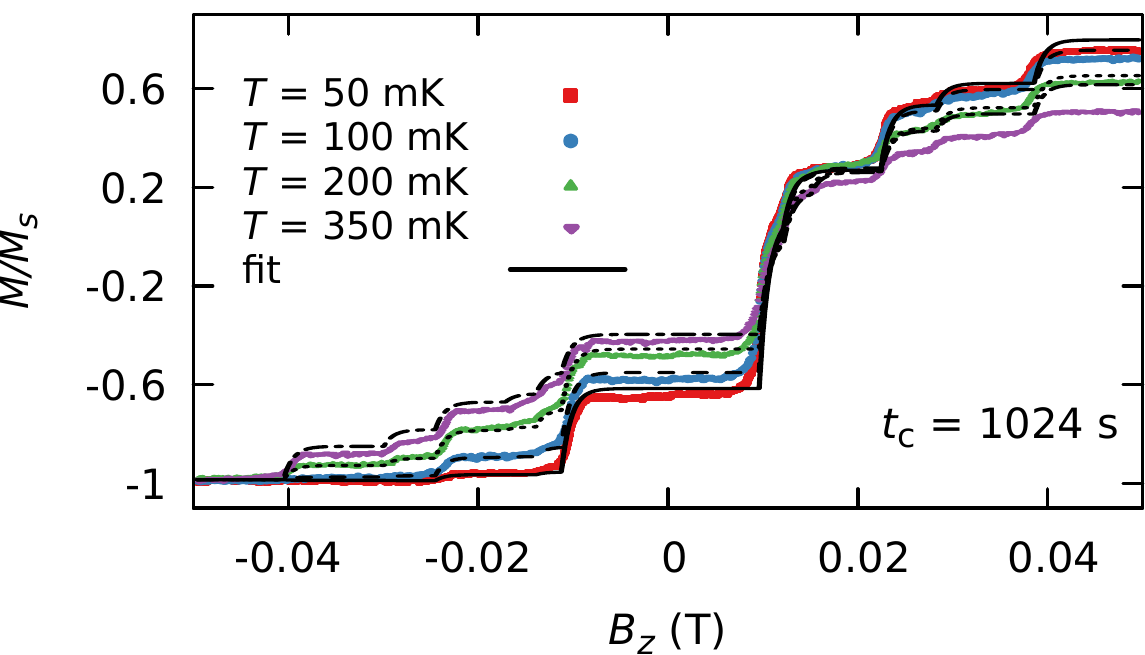}
    \end{minipage}
    \caption{Fit of the magnetization curves measured with $dB_z/dt=8$~mT/s and (\textit{top}) fixed $T=50$~mK for different cooling times, (\textit{bottom}) fixed $t_\text{c} = 1024$~s for different $T$.
    The fitting parameters are the initial populations of the hyperfine states and the relaxation is assumed to be dominated by incoherent quantum tunneling events occurring with the probability, $P_{\text{LZ}} = 0.5$. 
    }
	\label{fig:m_fit}
	\vspace{-15pt}
\end{figure}

As we continue to sweep the applied magnetic field, new steps that correspond to the subsequent level anticrossings are observed.
Their heights will no longer depend only on the initial population of the hyperfine states but also on the previous tunneling events.
One should notice that the small relaxation that is seen in-between steps is mainly due to dipolar interactions in the crystal (both of electronic and hyperfine origin) that broaden the resonance.
Besides QTM events that occur close to zero field, the hysteresis curve also displays a broad step at higher applied fields which is due to a direct spin-phonon relaxation process~\cite{urdampilleta2012molecular}.
This is a common characteristic of lanthanide SMMs which is the result of the unquenched orbital angular momentum.

We can now present the experimental protocol used to evidence and investigate the thermalization of $^{159}$Tb nuclear spins (see inset of Fig.~\ref{fig:therm_zd}a).
In order to start with a reproducible initial state we sweep back and forth through the zero field resonances until we reach a demagnetized state characterized by $ M  = 0$.
Through this procedure we heat-up the nuclear spin population to an effective temperature much higher than the cryostat temperature.
We then saturate the sample in a high longitudinal magnetic field ($B_z = -1.3$~T).
During this stage, we polarize the molecular spins, that is, all the molecules will be characterized by $m_J = +6$ while the nuclear spin population remains a out of equilibrium.
We keep the sample polarized in $B_z = -1.3$~T for a definite time that we call the \textit{cooling time} ($t_{\text{c}}$).
During this time the population of the $^{159}$Tb nuclear spins is allowed to evolve towards thermal equilibrium.
The last step is to read-out the nuclear spin states by inverting the applied field while measuring the $M(B_z)$ characteristic.
Figure~\ref{fig:therm_zd}a shows a zoom of the measured magnetization curves for increasingly larger cooling times.
We can see that the steps corresponding to excited hyperfine states, $(m_J = +6, m_I > -3/2)$ for $B_z < 0$~T, gradually diminish and then disappear as the system evolves towards thermal equilibrium.
Describing the above highlighted dynamics is the main focus of the current study.

\vfill
\noindent
\textbf{Theoretical analysis.}
We start the theoretical analysis by describing the numerical procedure by which we fit the magnetization curves and obtain the time dependence of the populations of the hyperfine states.
The principles underlying the analysis of $M(B_z)$ characteristics were already exposed when we investigated the Landau-Zener dynamics in TbPc$_2$ diluted crystals~\cite{taran2019decoherence}. 
In the aforementioned study we used the knowledge about the equilibrium Boltzmann distribution to fit the magnetization curve and infer the tunneling probability while in the current study we take advantage of the acquired understanding to obtain the time evolution of the population of the nuclear spin states.

When studying the field sweeping rate dependence of the tunneling probability ($P_{\text{LZ}}$), we found a behaviour qualitatively different from the predicted Landau-Zener dynamics of an isolated spin.
The observed tunneling relaxation was shown to be dominated by the environmentally induced decoherence with the central feature for the present study being the 
$P_{\text{LZ}} \xrightarrow~ 0.5$ limit observed at small sweeping rates (compared to $P_{\text{LZ}} \xrightarrow~ 1$ expected for the adiabatic limit of the Landau-Zener dynamics).

The model for $M(B_z)$ requires to keep track of the fractional populations of the hyperfine levels, $\Ket{m_J, m_I}$, at an applied longitudinal field $B_z$, which we denote by $n (m_J, m_I; B_z)$.
The initial state corresponds to the polarized sample in an applied field $B_z = -1.3$~T, thus, the only non-zero populations are the ones with $m_J = +6$. 
Note that, by definition, the fractional populations satisfy the relation: $ \sum_{m_J, m_I} n (m_J, m_I; B_z) = 1$.
With this notation, the normalized magnetization at an applied field $B_z$ is given by: 
\begin{equation}
    \label {eq:M}
    M(B_z)/M_\text{s} = \sum_{m_J, m_I} -\frac{m_J}{|m_J|}n(m_J, m_I; B_z)
\end {equation}
With $m_J = \pm 6$ and $m_I = -3/2 \dots 3/2$.
As we sweep the magnetic field, the change in the population of the hyperfine states is assumed to happen only at level anticrossings in the Zeeman diagram (Fig.~\ref{fig:therm_zd}b) with the relation between the tunneling probability and magnetization step given by: $P_{\text{LZ}} = |\Delta M / (2M_{\text{in}})|$.
Where $M_{\text{in}}$ is the initial magnetization of the states involved in the anticrossing and $\Delta M$ is the height of the relaxation step.
The fit of the magnetization curve (see Methods) for different  $t_{\text{c}}$ at fixed $T$ and fixed  $t_{\text{c}}$ with varying $T$ is shown in Fig.~\ref{fig:m_fit}. A good agreement with the experimental data is observed for $T < 300$~mK for all values of  $t_{\text{c}}$.

\begin{figure}
    \centering
    \begin{minipage}{0.5\textwidth}
        \includegraphics[width=\textwidth]{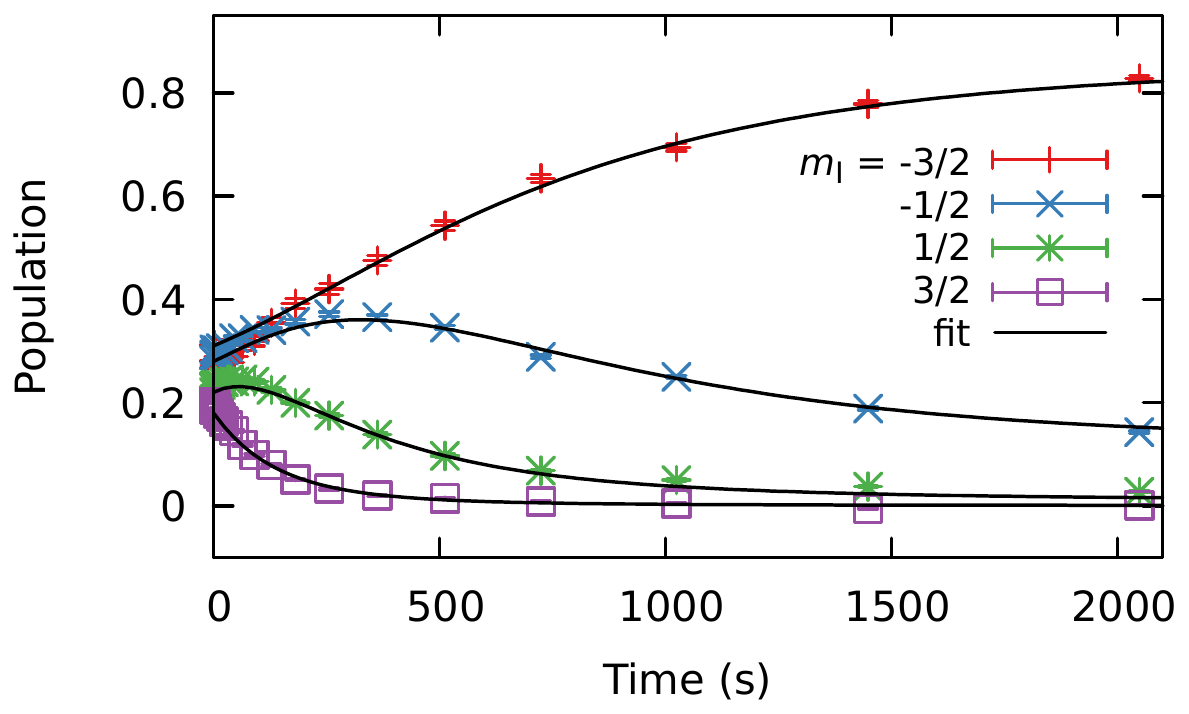}
    \end{minipage}
    \vfill
    \begin{minipage}{0.5\textwidth}
        \includegraphics[width=\textwidth]{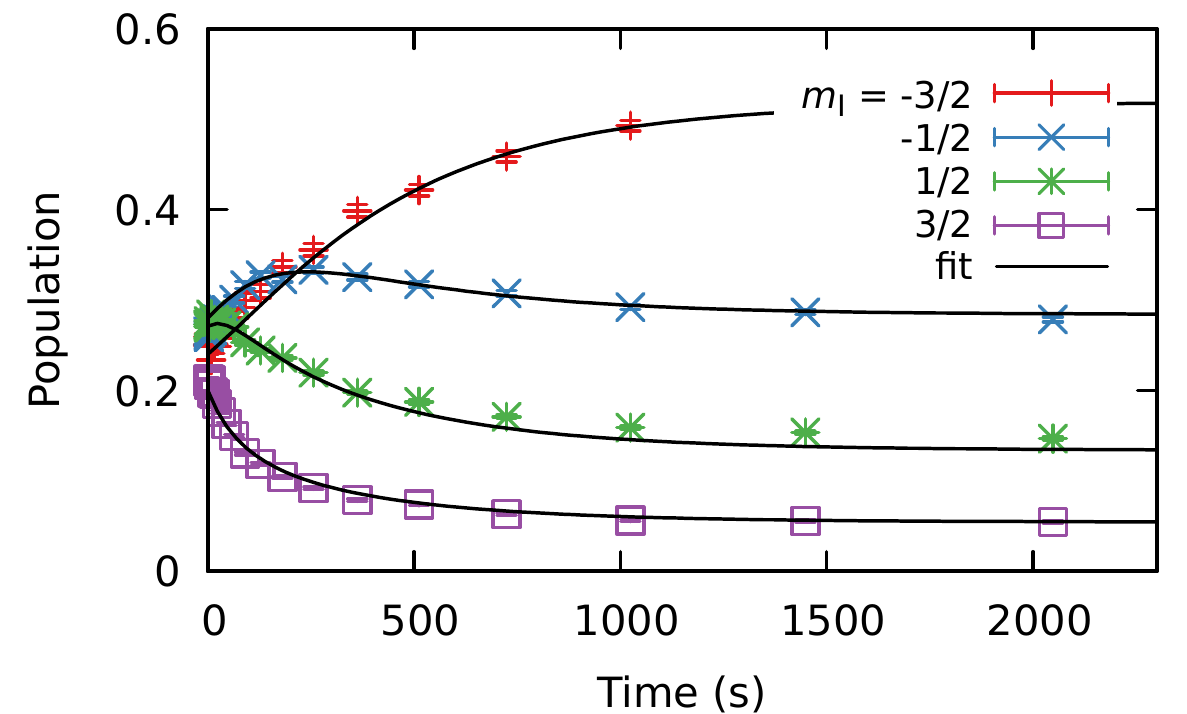}
    \end{minipage}
    \caption{
    Evolution towards thermal equilibrium of the populations of the hyperfine levels at (\textit{top}) 50~mK and (\textit{bottom}) 200~mK evaluated in the context of the master equation~(\ref{eq:master}). The fitting parameters are the de-excitation transition rates, $\gamma_{m}$.  
    }
	\label{fig:master}
	\vspace{-15pt}
\end{figure}

Before continuing with our analysis let's discuss the assumptions taken when constructing the above presented model for $M(B_z)$.
First, it is considered that the tunneling probability does not depend on the sample's magnetization, that is, all collective processes can be neglected.  
Due to dilution, TbPc$_2$ molecules are coupled only by weak dipolar interactions which are assumed to have the sole effect of giving a finite width (taken from experiment to be around 1~mT) to the relaxation steps.
Then, the characteristic time of the read-out technique (the time needed to sweep the $[-40:40]$~mT field range) which is around 10~s for the chosen sweeping rate of 8~mT/s, is much smaller than the characteristic relaxation times of the $^{159}$Tb nuclear spin.
This assumption is valid at low temperatures were the time needed to reach the thermal equilibrium is of the order of thousands of seconds (Fig.~\ref{fig:therm_zd} and Fig.~\ref{fig:master}) but starts to break down for temperatures larger than 300~mK (see Fig.~\ref{fig:m_fit}).
One has to remark that the rate of the electronic inter-well direct spin-phonon transitions also increases with temperature and constitutes another factor that limits the validity range of the presented technique.

Figure~\ref{fig:master} shows the evolution of the population of the hyperfine states as a function of $t_c$ obtained from the fit of the $M(B_z)$ characteristics.
We can see that the initial state corresponds indeed to a strongly non-equilibrium configuration, the populations of the hyperfine states being close to equal.
Also, at 50~mK, the relaxation is rather slow as the equilibrium Boltzmann distribution is reached on the time scale of thousands of seconds.
Another observation regards the population of $\ket{+6, -1/2}$ state (and to some degree of $\ket{+6, +1/2}$ which remains practically unchanged during the initial phase of the thermalization process.
This suggests that the relaxation process that brings the system to the equilibrium Boltzmann distribution follows the selection rule: $\Delta m_I = \pm 1$.

To get further insight into the relaxation process(es) that dominates the observed dynamics,
we model the thermalization of the nuclear spins in $B_z = -1.3$~T where the sample is polarized.  
For this,
we make use of a standard master equation for a memoryless, Markovian evolution:
\begin{equation}
\label {eq:master}
    \frac{d}{dt}n(m;t) = \sum_{q=m \pm 1} [\gamma_q^{m} n(q;t)-\gamma_{m}^q n(m; t)]
\end{equation}
Where $m$ and $q$ denote the hyperfine states and take values between $-3/2 \dots 3/2$ while  
$m_J = +6$ for all the states involved in the relaxation. 
$\gamma_q^m$ coefficients denote the transition rate from the state $\ket{+6, q}$ to the state $\ket{+6, m}$.  
The relaxation rates $\gamma_q^m$ obey the detailed balance condition: $\gamma_m^{m'}/\gamma_{m'}^m = \exp \left(\beta(E(m) - E(m'))\right)$, with $\beta = 1/(k_\text{B} T)$. 
The sum in~(\ref{eq:master}) is taken only over the nearest neighbour levels to reflect the above selection rule and reduce the number of fit parameters.
The fit of the master equation~(\ref{eq:master}) (see Methods) to the thermalization process at 50~mK is shown as black lines in Fig.~\ref{fig:master} with the three de-excitation rates, 
$\gamma_{m}^{m-1}$ (from now on denoted simply $\gamma_{m}$) as the only fitting parameters. 

\begin{figure}
    \centering
    \begin{minipage}{0.5\textwidth}
        \includegraphics[width=\textwidth]{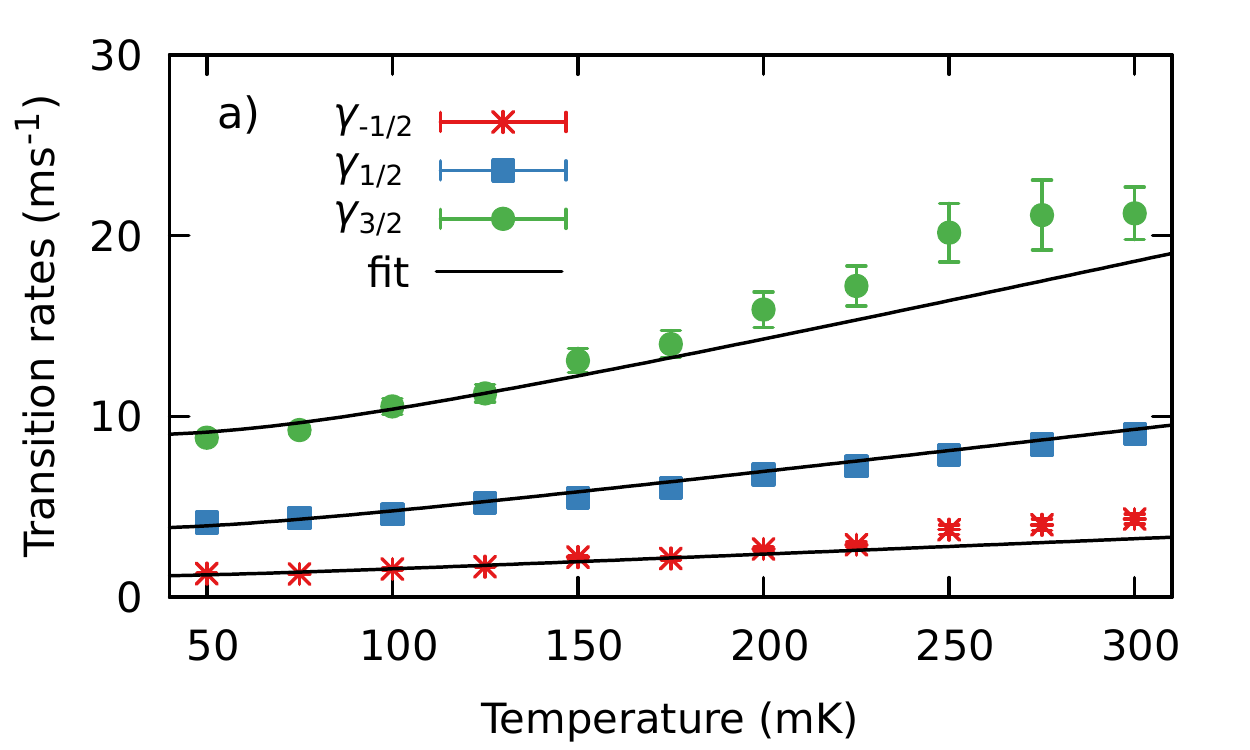}
    \end{minipage}
    \vfill
    \begin{minipage}{0.5\textwidth}
        \includegraphics[width=\textwidth]{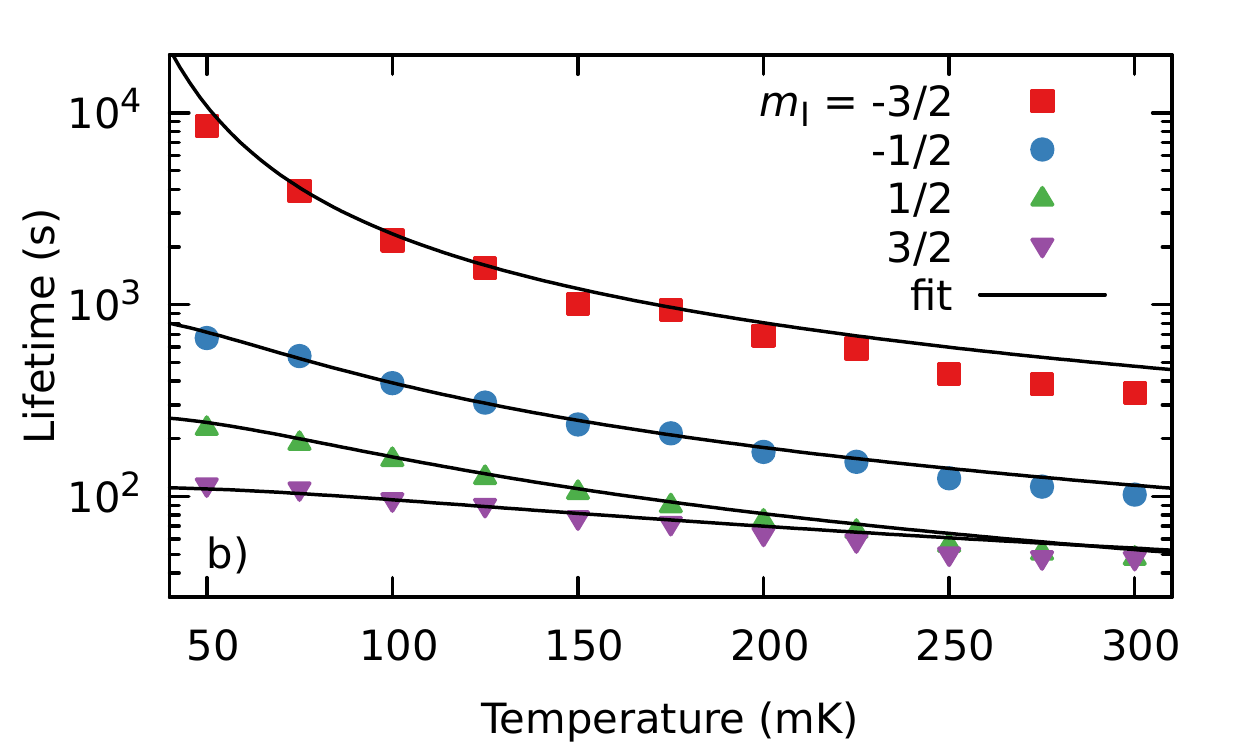}
    \end{minipage}
    \caption{(a) Temperature dependence of the relaxation rates fitted to a direct relaxation process characterized by a spontaneous and a induced component and given by Eq.~(\ref{eq:gamma}). 
    (b) Temperature dependence of the lifetime of the hyperfine levels computed by using: $1/\tau_{m} = \left(\gamma_{m}^{m-1} +  \gamma_{m}^{m+1}\right)$.
The continuous lines denote the evaluation of the level lifetimes using Eq.~(\ref{eq:gamma}).
    In the $T \xrightarrow~0$ limit, the lifetime of the ground state, $\ket{+6, -3/2}$, becomes infinite, while $\tau_m$ of the excited states are mainly determined by the spontaneous emission process.}
	\label{fig:gamma_t}
	\vspace{-15pt}
\end{figure}

In order to identify the relaxation process we repeat the above presented analysis for temperatures up to 300~mK where the model for the magnetization curve starts to break down and the estimation of the population of the hyperfine states is no longer accurate.
Figure~\ref{fig:gamma_t}a shows the obtained temperature dependence of the relaxation rates.
The transition rates increase with the spacing between the hyperfine levels.
Also, for temperatures roughly smaller than 100~mK, the relaxation process becomes temperature independent.
This suggests that the transition rates are determined by the sum of a spontaneous and an induced process.
Thus, considering a pair of adjacent hyperfine levels $\ket{+6, m}$ and $\ket{+6, m-1}$ separated in energy by $\Delta E_{m}$, the transition rate $\gamma_{m}$ can be expressed as:
\begin{equation}
\label {eq:gamma}
    \gamma_{m} = 
    {\cal F}(\Delta E_{m})\frac{\exp\left(\beta \Delta E_{m} \right)}{\exp\left(\beta \Delta E_{m} \right)-1}
\end{equation}

The fit curves are shown in Fig.~\ref{fig:gamma_t}a as black lines with ${\cal F}(\Delta E_{m}) = 1.09$, $3.81 \text{ and } 8.86$ ($\times 10^{-3}$s$^{-1}$) for the three de-excitation transitions, $\gamma_m$, with $m = -1/2$, $1/2$ and $3/2$.
The expression~(\ref{eq:gamma}) works especially well at low temperatures ($T \lesssim 200$~mK) while the deviations that we start to see at higher temperatures suggest that the inclusion of higher order processes (\textit{e.g.} Raman or Orbach mechanisms~\cite{abragam1970}) may play a role in the relaxation dynamics. 

We can also compute the lifetime, $\tau_{m}$, of each hyperfine level through the expression:  
$1/\tau_{m} = \left(\gamma_{m}^{m-1} +  \gamma_{m}^{m+1}\right)$.
Figure~\ref{fig:gamma_t}b shows the evaluation of the lifetime of the hyperfine levels using the determined relaxation rates from the master equation and also by using Eq.~(\ref{eq:gamma}) (black lines in Fig.~\ref{fig:gamma_t}b).
As one can expect, in the $T \xrightarrow~0$ limit, the lifetime of the ground state, $\ket {+6, -3/2}$ becomes infinite, as there are no phonons with sufficient energy to excite the nuclear spin, while the finite lifetime of the excited states are determined by the spontaneous emission process.

In order to interpret ${\cal F}(\Delta E_{m})$, 
the coupling mechanism between the nuclear spins and 
phonon bath should be considered.
The electronic shell of the Tb$^{3+}$ ion couples to 
the lattice vibrations through 
the ligand field interaction,
while the link between the electronic configuration and the $^{159}$Tb nuclear spins is made through 
the hyperfine (mainly the spin orbit term) and quadrupolar interactions.
Thus, we follow Ref.~\cite{gatteschi2006molecular,leuenberger2000spin} and compute 
the transition rates induced by 
the phonon modulation of the nuclear spin Hamiltonian and obtain: 
\begin{equation}
\label {eq:F}
    {\cal F}(\Delta E_{m}) = 
    \frac{m_J^2 \delta A_{\text{hyp}}^2 \Delta E_{m}^3}{6 \pi \hbar^4 \rho c^5} (I(I+1) - m(m-1))
\end {equation}
Where $\rho$ is the crystal density and $c$ is the sound velocity.
To our knowledge, measurements of both $\rho$ and $c$ in TbPc$_2$ crystals were not reported so far, thus one has to take them as free parameters.
Especially, variations in the sound velocity will have a big impact on the relaxation rate as it enters in Eq.~\ref{eq:F} to the fifth power.
Also, the Debye model (a linear dispersion for the acoustic phonon modes) used in the derivation of Equation~(\ref{eq:F}) is an over-simplification.
However, as we miss the value of $\rho$ and the details on the lattice modes, we will be content with a rough estimation of the order of magnitude of the relaxation rates.
Thus, by using Eq.~(\ref{eq:gamma}) and ~(\ref{eq:F}) for the evaluation of $\gamma_{-1/2}$ at 50~mK we get: $\rho c^5 \approx 3.5 \times 10^{19} \text{ kg}\cdot\text{m}^2/\text{s}^5$.
And by setting a sensible value for $\rho = 1500 \text{ kg}/\text{m}^3$, we obtain $c = 1877 \text{ m}/\text{s}$, which is a reasonable enough value (for example, $c = 1450 \text{ m}/\text{s}$ was used to explain phonon-assisted tunneling in Mn$_{12}$-ac~\cite{leuenberger2000spin}) to confirm the proposed mechanism for the thermalization process.

\vfill
\noindent
\textbf {Discussion}\\
The identified direct relaxation channel between 
the $^{159}$Tb nuclear spins and the phonon bath          
is a rather unexpected find for 
the relaxation of a nuclear spin embedded in a
molecular complex.
For example,     
measurements on $^{55}$Mn nuclear spins at the core of Mn$_{12}$-ac SMM    
were used to check    
the predictions of the spins bath theory for     
the dynamics of \textit{molecular spin--nuclear bath} coupled system\cite{morello2007dynamics}.    
Most of the observed phenomenologies, 
with the exception of $^{55}$Mn thermalization,
were successfully explained.   
It was suggested that  
$^{55}$Mn nuclear spins thermalize through 
the quantum dynamics of the molecular spins because
the spin lattice interactions  
were found inefficient to explain  
the measured relaxation rates.
However, so far no theoretical solution to this problem
was found.
The relaxation mechanism that 
we evidence for
$^{159}$Tb is not efficient in the case of 
$^{55}$Mn nuclear spins because
the hyperfine interaction in
transition metal ion compounds is 
around one order of magnitude smaller.

Another interesting example to consider is 
the spin lattice relaxation of $^{159}$Tb in
TbPc$_2$ molecular spin transistor geometry\cite{thiele2013electrical}.
The relaxation process, 
with a characteristic time of tens of seconds, 
was found to be dominated by 
the interaction with the electrons 
that tunnel through the molecular quantum dot. 
The comparison between the two experiments, that 
share the same molecular complex placed in 
very different environments, suggests that
the direct relaxation mechanism that
we highlight in this work
sets the lower limit for the 
nuclear relaxation rate in
potential lanthanide SMMs based spintronics devices.

Recently, the effect of the nuclear isotopes on
the molecular spin relaxation was evidenced 
when comparing two isotopologues lanthanide dimers
~\cite{moreno2019quantum}.
It was shown that the presence of the nuclear spin 
leads to
a significant increase of the relaxation rate at
crossover temperatures, that is, when 
molecular spin tunneling and phonon assisted transitions 
occur with comparable rate.
We suggest that the missing ingredient for 
constructing a quantitative explanation of the observed dynamics is
the thermal fluctuations of the nuclear spin.

In conclusion, we have investigated 
the thermalization of $^{159}$Tb nuclear spin belonging to 
the archetypical TbPc$_2$ complex and proven that 
the relaxation is due to 
the phonon modulation of the hyperfine interaction.
The uncharacteristic, sub-kelvin, phonon induced hyperfine fluctuations should be especially important in 
the crossover temperature domain. 
Through this work, we try to argue that 
the direct contact of the nuclear spins to 
the phonon modes in lanthanide compounds is 
an important feature that has to be considered both in 
the continuous search for molecular compounds with optimized magnetic properties and 
fundamental investigations on the spin bath dynamics. 

\vfill
\noindent
\textbf {Methods}\\
\footnotesize 
\textbf{$\upmu$SQUID measurements:}
$\upmu$SQUID measurements are performed on micrometer sized monocrystals containing TbPc$_2$ diluted in YPc$_2$ matrix with [Tb/Y] ratio of 1\%. The synthesis and detailed characterization of the sample is given in Ref.~\cite{konami1989analysis}.
The crystals are measured with an array of $\upmu$SQUIDs placed in a home built $^3$He/$^4$He dilution refrigerator.
The field can be applied in any direction by using a three orthogonal coil system.
The easy axis of the molecules was identified by using the transverse field method.

\noindent
\textbf {Numerical fit:}
The fit of the magnetization curves to the Eq.~\ref{eq:M} and the fit of the thermalization dynamics to the master equation given by
Eq.~\ref{eq:master} was done using the following procedure:\\
- solve the models numerically at the experimental points.\\
- minimize the least square deviation by using a nonlinear least-squares (NLLS) Marquardt-Levenberg algorithm.\\
The second step is done with the use of \textit{gnuplot}.

\vfill
\noindent
\textbf {Acknowledgements} \\
We acknowledge the Alexander von Humboldt Foundation and the ERC advanced grant MoQuOS No. 741276.

\vfill
\noindent
\textbf {Author contributions} \\
All the authors participated in designing, conducting, analysing the experiment and in writing the paper.

\vfill
\noindent
\textbf {Competing interest} \\
The authors declare no competing interest.

\bibliography{bib}

\end{document}